\documentclass{ws-p8-50x6-00}
\usepackage{amssymb, amsmath}
\begin{document}
\def\lsim{\:\raisebox{-0.5ex}{$\stackrel{\textstyle<}{\sim}$}\:}
\def\gsim{\:\raisebox{-0.5ex}{$\stackrel{\textstyle>}{\sim}$}\:}
\def\sh{\mbox{\texttt h}}
\def\et{\mbox{${E\!\!\!/}_T$}} 
\def\rpv{\mbox{${R\!\!\!/}_p$}} 
\def\eplem{\mbox{$e^+e^-$}}
\def\NCA{\em Nuovo Cimento}
\def\NIM{\em Nucl. Instrum. Methods}
\def\NIMA{{\em Nucl. Instrum. Methods} A}
\def\NPB{{\em Nucl. Phys.} B}
\def\PLB{{\em Phys. Lett.}  B}
\def\PR{\em Phys. Reports}
\def\EUJ{{\em Eur. Phys. J.} C}
\def\PRL{\em Phys. Rev. Lett.}
\def\PRD{{\em Phys. Rev.} D}
\def\ZPC{{\em Z. Phys.} C}
\def\PR{{\em Phys. Reports}}
\def\RMP{\em Rev. of Modern Physics}
\def\JHEP{\em Journal of High Energy Physics}
\def\pramana{\em Pramana}
\title{Physics potential of the LHC}

\author{Rohini M. Godbole}

\address{Center for Theoretical Studies, Indian Institute of Science,
Bangalore 560 012,
INDIA\\E-mail: rohini@cts.iisc.ernet.in}

\begin{flushright}
IISc-CTS/18/00 \\
hep-ph/0011237 
\end{flushright}

\vskip 25pt
\begin{center}

{\large\bf Physics potential of the LHC\footnote{Invited talk presented at the 
joint session of Asia Pacific Physics Conference  2000 and III ACFA
 Linear Collider Workshop, August 2000, Taipei, Taiwan.}}    
       \\
\vskip 25pt

{\bf                        Rohini M. Godbole } \\ 

{\footnotesize\rm 
                      Centre for Theoretical Studies, 
                     Indian Institute of Science, Bangalore 560 012, India. \\ 
                     E-mail: rohini@cts.iisc.ernet.in  } \\ 

\vskip 20pt

{\bf                             Abstract 
}

\end{center}

\begin{quotation}
\noindent
The basic aim of physics studies at the LHC is to unravel the mechanism
responsible for the spontaneous symmetry breaking in the Standard Model (SM).
 In the currently accepted theoretical picture, this translates into finding 
`direct' experimental evidence for the Higgs sector. TeV scale supersymmetry
(SUSY)  provides a very 
attractive solution to the `naturalness' problem that theories with 
elementary scalar fields have. Hence in this talk I will  summarise the
physics potential of the LHC for searching for Higgs and Supersymmetry as well
as for measurement of the parameters of the Higgs sector and the SUSY model. 
Theories with localised gravity (and large extra dimensions) give a credible 
option to have  Standard Model without the attendant `naturalness'
problems.  I will therefore also summarise the potential of LHC to probe 
these `large' extra dimensions.
\end{quotation}

\maketitle

\abstracts{ 
The basic aim of physics studies at the LHC is to unravel the mechanism
responsible for the spontaneous symmetry breaking in the Standard Model (SM).
 In the currently accepted theoretical picture, this translates into finding 
`direct' experimental evidence for the Higgs sector. TeV scale supersymmetry
(SUSY)  provides a very 
attractive solution to the `naturalness' problem that theories with 
elementary scalar fields have. Hence in this talk I will  summarise the
physics potential of the LHC for searching for Higgs and Supersymmetry as well
as for measurement of the parameters of the Higgs sector and the SUSY model. 
Theories with localised gravity (and large extra dimensions) give a credible 
option to have  Standard Model without the attendant `naturalness'
problems.  I will therefore also summarise the potential of LHC to probe 
these `large' extra dimensions.
}

\section{Introduction}

 LHC is the pp collider which is expected to start operation in year 
$\sim 2005$.  For the first three years we expect to collect $10 fb^{-1}$ 
luminosity per year whereas at the end of four years the integrated 
luminosity will be $100  fb^{-1}$ and an integrated luminosity 
of $300  fb^{-1}$ will be available only by the year 2011. This should be 
kept in mind while assessing the physics reach of LHC in various channels. 
LHC will be a factory of all kinds of particles  W/Z, t, b etc. 
\begin{table}[t]
\caption{\em Total number of events for different types of final states 
expected at LHC per year for $10  fb^{-1}$ integrated luminosity.\label{tab:tab1}}
\begin{center}
\footnotesize
\begin{tabular}{|c|c|c|c|}
\hline
Process&$\#$ of events/yr.& Process&$\#$ of events/yr.\\
\hline
&&&\\
$W \rightarrow e \nu$ & $10^8$ & $Z \rightarrow e^+ e^-  $ & $10^{7}$\\
&&&\\
$ t \bar t $ & $ 10^7 $ & $b \bar b$ & $ 10^{12}$\\
&&&\\
$\tilde g \tilde g$ (TeV) & $10^4$ & $jets ( > 200\;{\rm GeV}$) & $10^9 $\\
\hline
\end{tabular}
\end{center}
\end{table}
Table~\ref{tab:tab1} gives the expected number of events/year for 
$ 10 fb^{-1}$ luminosity for different final states. This 
will provide a laboratory to make precision
studies of EW theory, QCD, flavour physics, CP violation etc. and will
further our theoretical understanding of the workings of the Standard Model
(SM) and beyond.
In all these studies the primary goal of the LHC experiments will be to
understand the `physics of the Spontaneous Symmetry breaking of EW 
symmetry'.  There are two major components to the physics studies at 
LHC (i) its discovery potential and (ii) potential to do precision 
studies. In this talk I will concentrate on the aspects of both these 
studies that have a bearing on the abovestated primary  goal. Thus I 
will try to summarize the ability of LHC to throw light on the mechanism 
of EW symmetry breaking.  To this end I will start with the case of 
SM Higgs where the new results to be reported are investigations into
ability of LHC to measure properties of the Higgs to nail it down as {\bf
the} SM
Higgs or otherwise. Then I will discuss the case of supersymmetric Higgs
where I will concentrate on effects of light supersymmetric particles on the
search of light neutral higgs of the MSSM. I will list some of the new
developments in SUSY phenomenology which focus on the possibility of
determining the model parameters once SUSY is found. I will end by
discussing the reach of  LHC to search for large extra dimensions which
is an example of new physics which might obviate the need for TeV scale
supersymmetry in order to  keep the Higgs scalar light.

\section{ Identification of benchmarks to gauge LHC potential}
\begin{figure}[htb]                
       \centerline{
      \includegraphics*[scale=0.53]{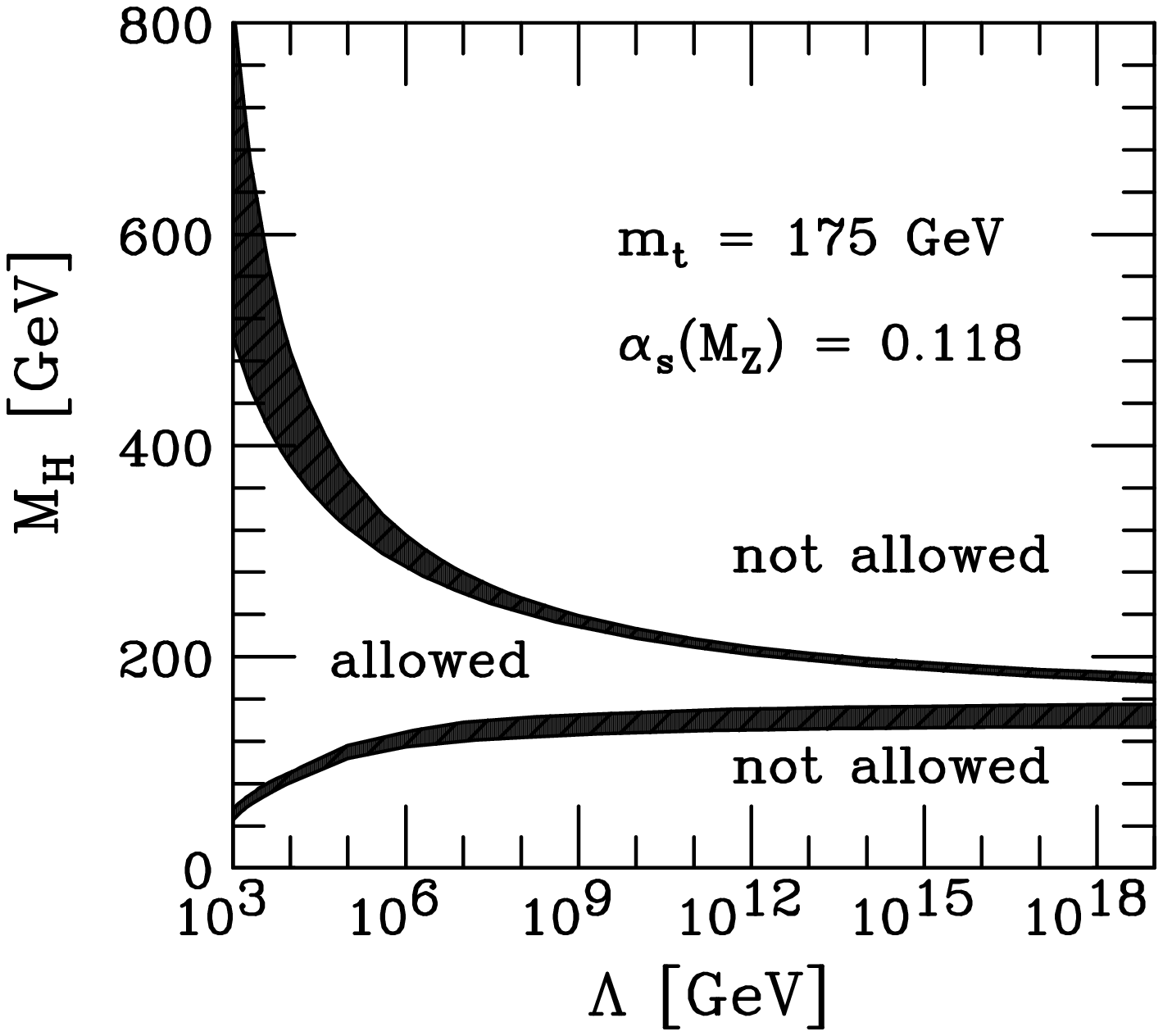}
      \includegraphics*[scale=0.35]{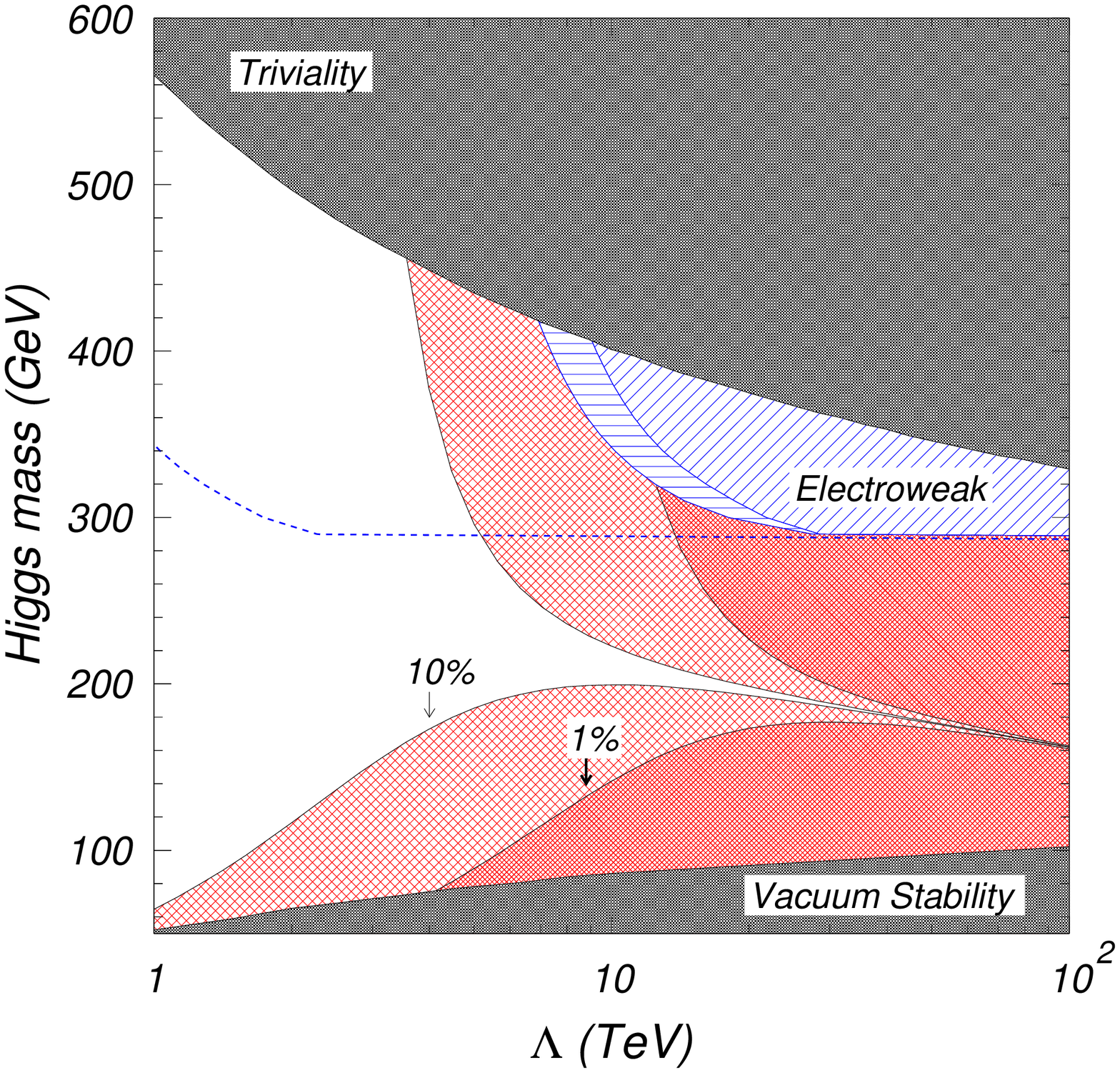}
  }
\caption{\em  Limits on the Higgs mass in the SM and 
beyond\protect\cite{hambye,murayama}.
\label{fzero}}           
\end{figure}

The precision measurements from LEP I as well as LEP-200 data on
$W^+W^-$ production has proved that the SM is the correct theory of 
EW interactions, at least as an effective interaction and that the 
higgs is light. Precision data give a limit $m_h \lesssim 210 $  GeV     
at $95 \%$  level\cite{Komamiya} and at the same time direct
searches give  $m_h > 113.3$ GeV.
On the theoretical side there exist precise predictions for the couplings
of the higgs scalar but for the mass there exist only limits. These come
from different theoretical considerations of vacuum stability as well as
triviality of the $\phi^4$ theory. The figure in the left panel in  
fig.~\ref{fzero} taken from Ref.[2] shows the expected limits in
the SM as a funtion of the scale $\Lambda$, upto which SM should be the correct
description. The lesson to learn from this figure is $m_h \simeq 160 \pm  20$
GeV if SM is the correct description all the way upto the Plank scale.
The consistency of the above theoretical and experimental statements
implies that to nail down the Higgs mechanism as `the' mechanism of EW
symmetry breakdown,
\begin{description}
\item[I)] LHC should find `direct' experimental evidence for a light Higgs
particle with couplings as predicted in the SM,
\item[II)] LHC should find TeV scale supersymmetry (SUSY) if this is the right
solution of the hierarchy problem \footnote {Note here that SUSY is to be
considered as an example of the new physics at the TeV scale. However, this
is the only example that is well-defined and has specific predictions.}. 
Supersymmetry implies an upper limit on SUSY Higgs mass of about 135 (200)
GeV\cite{heinemeyer,quiros}, the larger number  is in a general framework
extending the MSSM.
\end{description}
However, recently more general analyses of correlations of the scale of
new physics and the mass of the Higgs have started\cite{barbie,murayama}. In
this analysis the assumption is that the SM is only an effective theory and
additional higher dimensional operators can be added. Based on very general
assumptions about the coefficients of these higher dimensional operators, an
analysis of the precision data from LEP-I with their contribution,  
alongwith a requirement that the radiative corrections to the $m_h$ do not
destabilize it more than a few percent, allows different regions in $m_h -
\Lambda$ plane. This is shown in the right panel in 
fig.~\ref{fzero}.  The lesson to learn from this figure taken from Ref. [6] is  
that a light higgs with $m_h < 130$ GeV will imply existence of new 
physics at the scale
$\Lambda < 2-3 $ TeV, whereas  $195 < m_h < 215$ GeV would imply 
$\Lambda_{Np} < 10$ TeV. These arguments, specialized to the case of 
Supersymmetry, do imply that SUSY ought to be at TeV scale to be
relevant to solve the fine tuning problem.
The above discussion suggests yet another need that LHC should
fulfil, {\it viz.}, 
\begin{description}
\item[III)] LHC should be able to look for new physics at a scale 
from 1 - 10 TeV
which is implied by demanding theoretical consistency of SM as a field
theory with a light Higgs.
\item[IV)] Should LHC not find a light scalar, it essentially 
would imply that the Higgs is not a fundamental scalar. Then LHC 
should be able to look for the `strongly interacting W' sector.
\item[V)] The idea of new extra dimensions\cite{ADD,RS} can obviate the
`naturalness' problem even with a `light' higgs. In this case LHC should be
able to look for evidence for extra dimensions.
\end{description}
So the potential of LHC has to be evaluated in view of how it can reach the
aims listed as I-V above.

\section { Search for the SM higgs at the LHC }

   The current limit on $m_h$ from precision measurements at LEP is 
$m_h < 210$ GeV at $95\%$ C.L. and limit from direct searches is 
$m_h\lesssim 113$ GeV\cite{Komamiya}. Tevatron is likely to be able to 
give indications  of the existence of a SM higgs,  by combining data in  
different channels together,\cite{bhat} for $m_h \lesssim 120$ GeV if 
Tevatron run II can accumulate $30  fb^{-1}$ by 2005.
The best mode for the detection of Higgs depends really on its mass. Due to
the large value of $m_t$ and the large $gg$ flux at LHC, the highest production
cross-section is via $gg$ fusion. 
\begin{figure}[htb]
\begin{center}
\includegraphics*[scale=0.3]{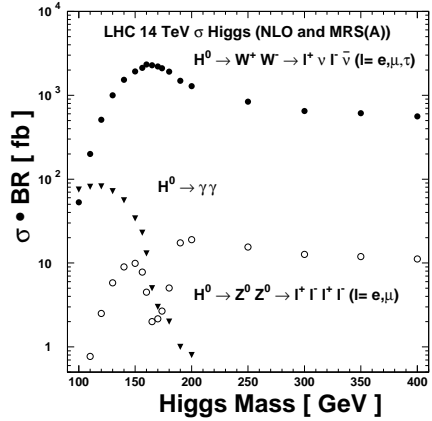}
\caption{\em Expected $\sigma \cdot BR$ for different detectable SM Higgs
decay modes~\protect\cite{michael}.}
\label{fone}
\end{center}
\end{figure}
Fig.~\ref{fone}\cite{michael} shows 
$\sigma \cdot BR$ for the SM higgs for various final states. 
The  search prospects are optimised by exploring different channels
in different mass regions.  The inclusive channel using $\gamma \gamma$ 
final states corresponds to  $\sigma \cdot  BR \; {\rm of\;\; only} \:  50 \: fb$,
but due to the low background it constitutes the cleanest channel for 
$m_h < 150$ GeV.
The important detector requirement for this measurement is good resolution
for $\gamma \gamma$ invariant mass. ATLAS should be able to achieve 
$\sim 1.3$ GeV and CMS $\sim 0.7$ GeV. The new developments in this 
case is a NLLO calculation for $\gamma \gamma$ background which is 
now available\cite{Binoth}. A much more interesting channel is 
production of a higgs recoiling against a jet in the process
$ gg \to h + {\rm jets} \to \gamma \gamma + {\rm jets}.$
The signal is much lower but is also much cleaner. The background in the
channel is also known\cite{kunszt}. The viability of this channel was
studied initially using only a tree level calculation\cite{Abdullin}. The
effect of resummation on the signal size has been recently
discussed\cite{Djouadi}. Use of this channel gives a significance of $\sim 5$
already at $30  fb^{-1}$ for the mass range $110 < m_h < 135 $ GeV.
A more detailed study of the channel $  pp \to h + t \bar t \to \gamma \gamma +
t \bar t$ with semileptonic decays of the $t/\bar t$ is now underway. This
is important also for the measurement of the $h t \bar t$ couplings.

For larger masses $(m_h \gsim 130$  GeV ) the channel 
$gg \to h \to Z Z^{(*)} \to 4l$ is the best channel.  
Fig.~\ref{fone} shows that, in the range $150 \: {\rm GeV}  < m_h < 190 $ 
GeV this clean channel, however, has a rather low $(\sigma \cdot  BR)$.  
The viability of $p \bar p  \to W W^{(*)} \to l \bar \nu \bar l \nu$
in this channel has  been demonstrated.\cite{dreiner} 
Thus to summarise for $m_h \lsim 180 $ GeV, there exist 
a large number of complementary channels whereas beyond that the 
gold plated  $4 l$ channel is the obvious choice. If the Higgs is
heavier, the  event rate will be too small in this channel 
({\it cf.} Fig.~\ref{fone}). Then the best option is to tag the forward jets
by studying the production of the Higgs in the process $ p \bar p \to W W q \bar q \to q \bar q h$.
\begin{figure}[htb]                
       \centerline{
             \includegraphics*[scale=0.38]{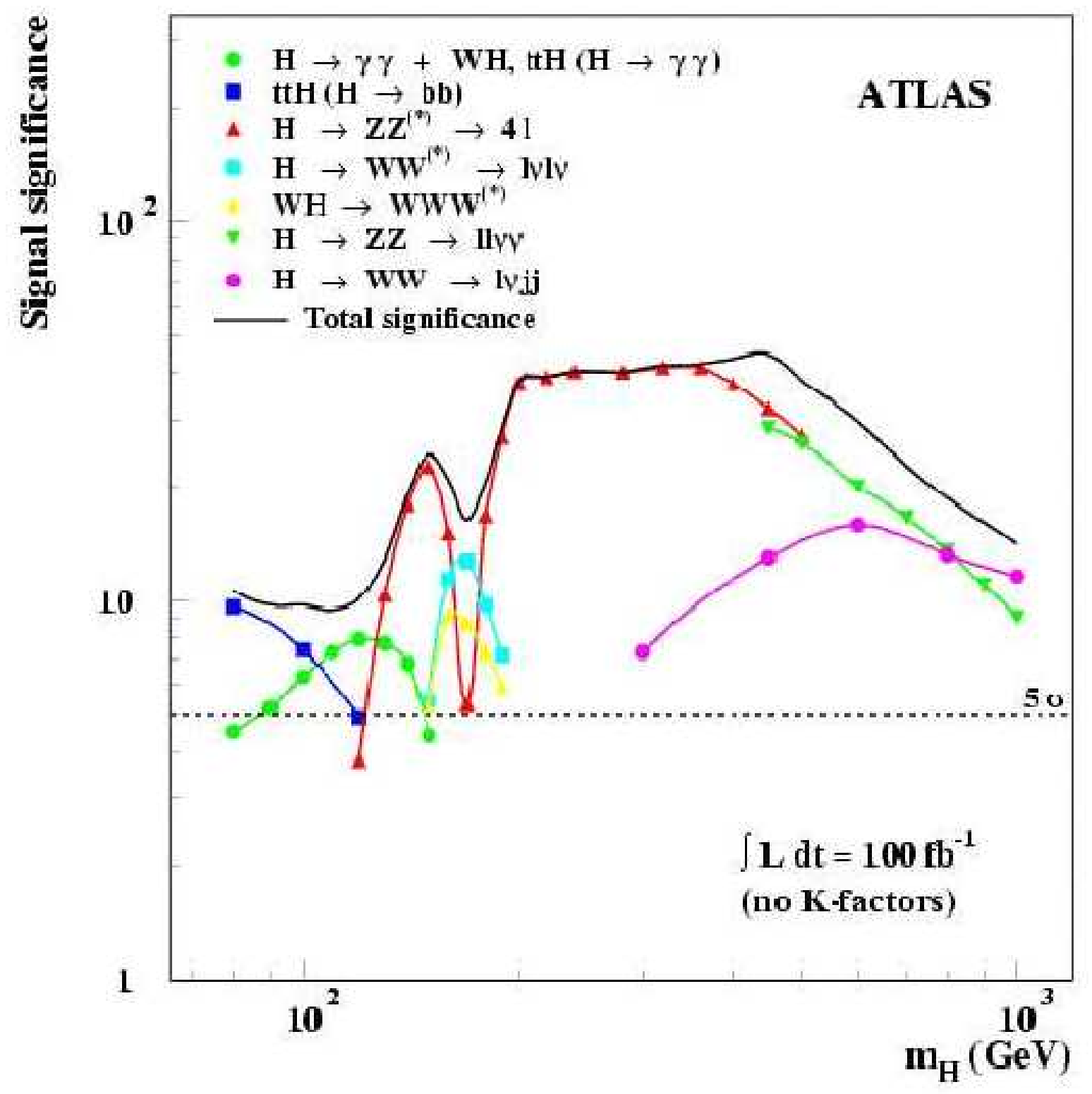}
              \includegraphics*[scale=0.45]{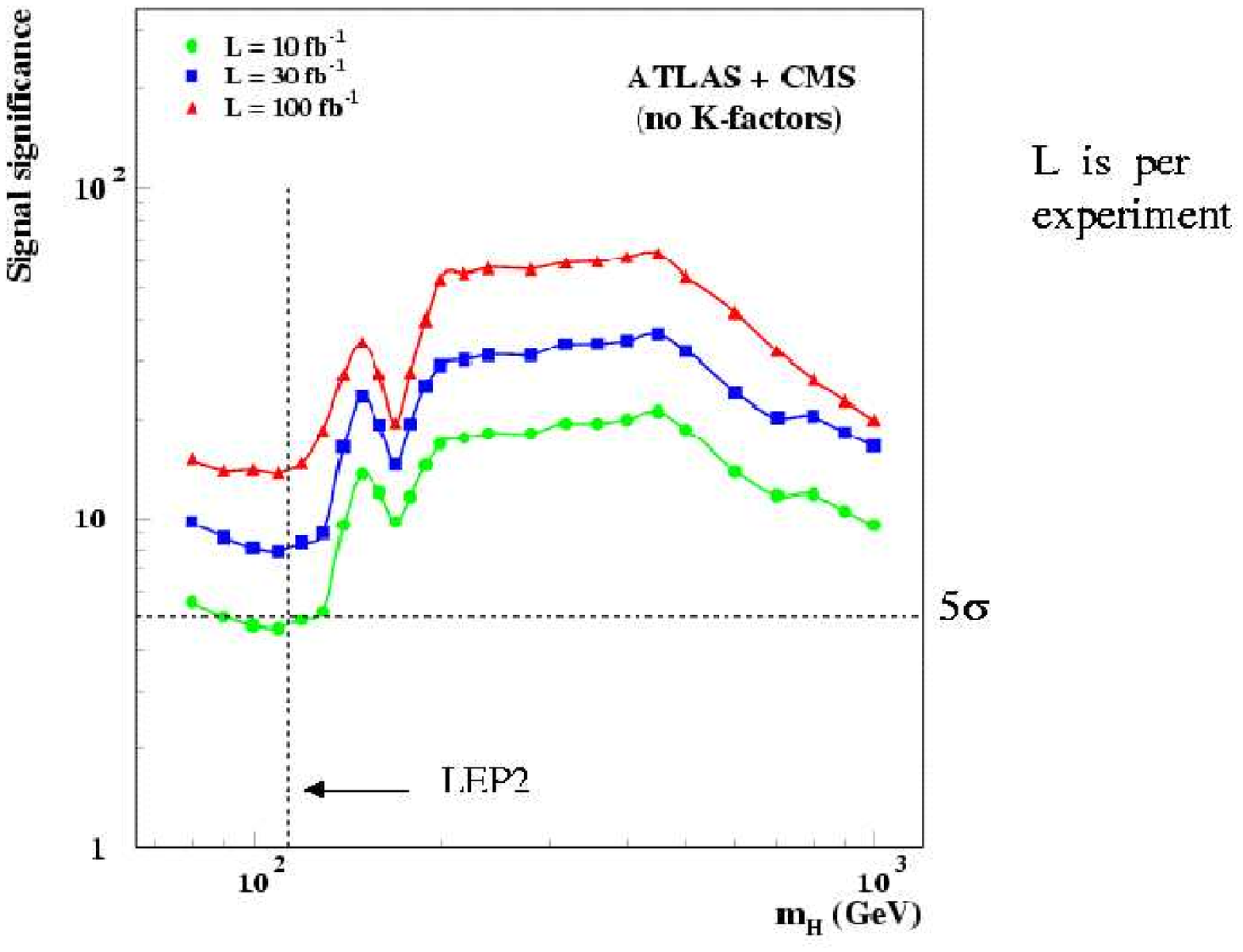}
                   }
        \caption{\em The expected significance level of the SM Higgs
signal at LHC~\protect\cite{fabiola}.\label{ftwo}}           
\end{figure}
The figure in the left panel in fig.~\ref{ftwo} shows the overall discovery 
potential of the SM
higgs in all these various channels whereas the one on the right shows the same
overall profile of the significance for discovery of the SM higgs, for three
different luminosities, combining the data that both ATLAS and CMS
can obtain. The figure shows that the SM higgs boson can be discovered
(i.e. signal significance $\gsim 5$) after about one year of operation 
even if $m_h \lsim 150$ GeV. Also at the end of the year the SM higgs boson 
can be ruled out over the entire mass range implied in the SM discussed earlier.

A combined study by LHC and ATLAS shows that a measurement of $m_h$ at $0.1\%$
level is possible for $m_h \lsim 500$ GeV, at the end of three years of high 
luminosity run. As the panel on the left in 
\begin{figure}[htb]
        \centerline{
             \includegraphics*[scale=0.40]{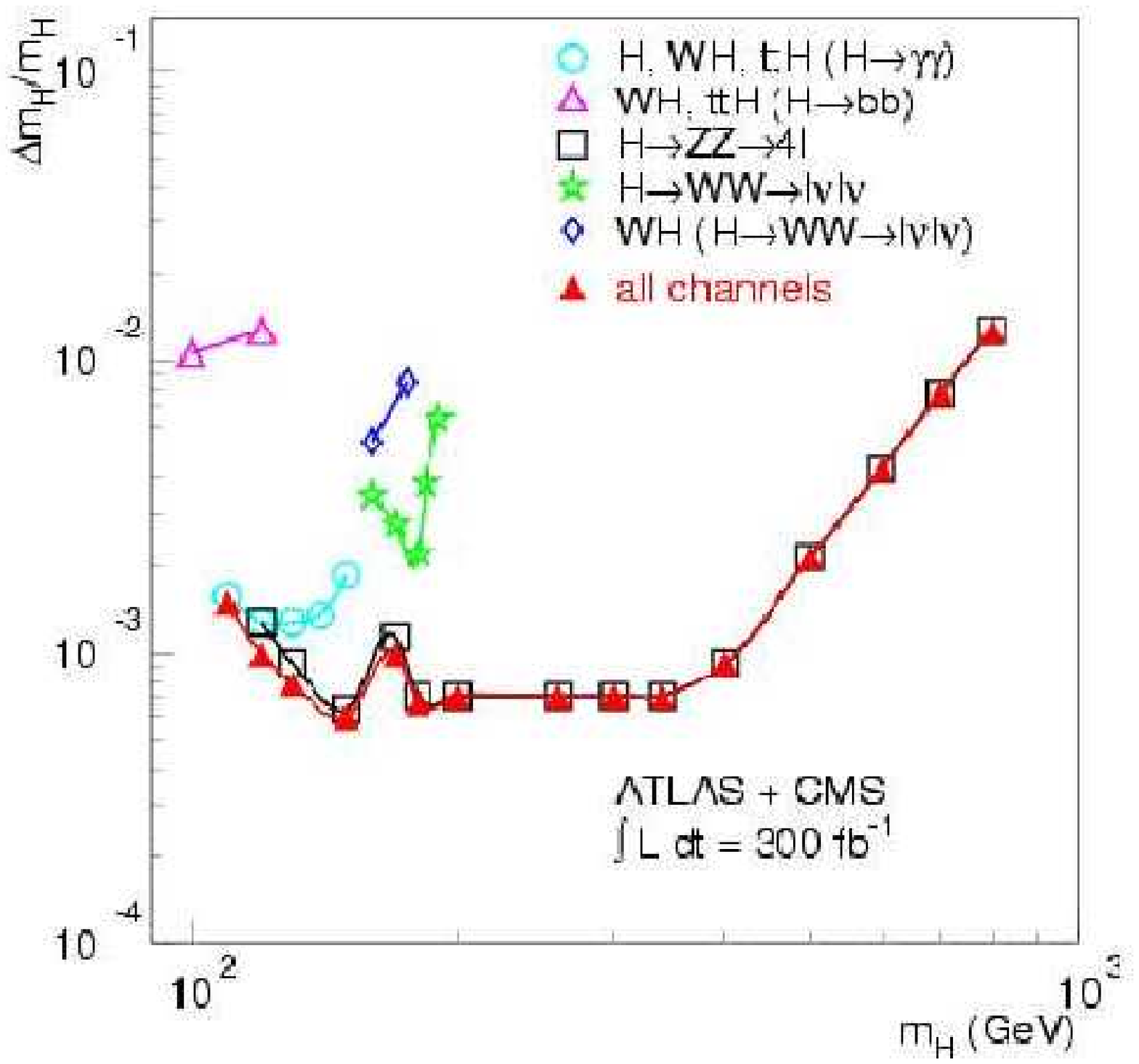}
              \includegraphics*[scale=0.38]{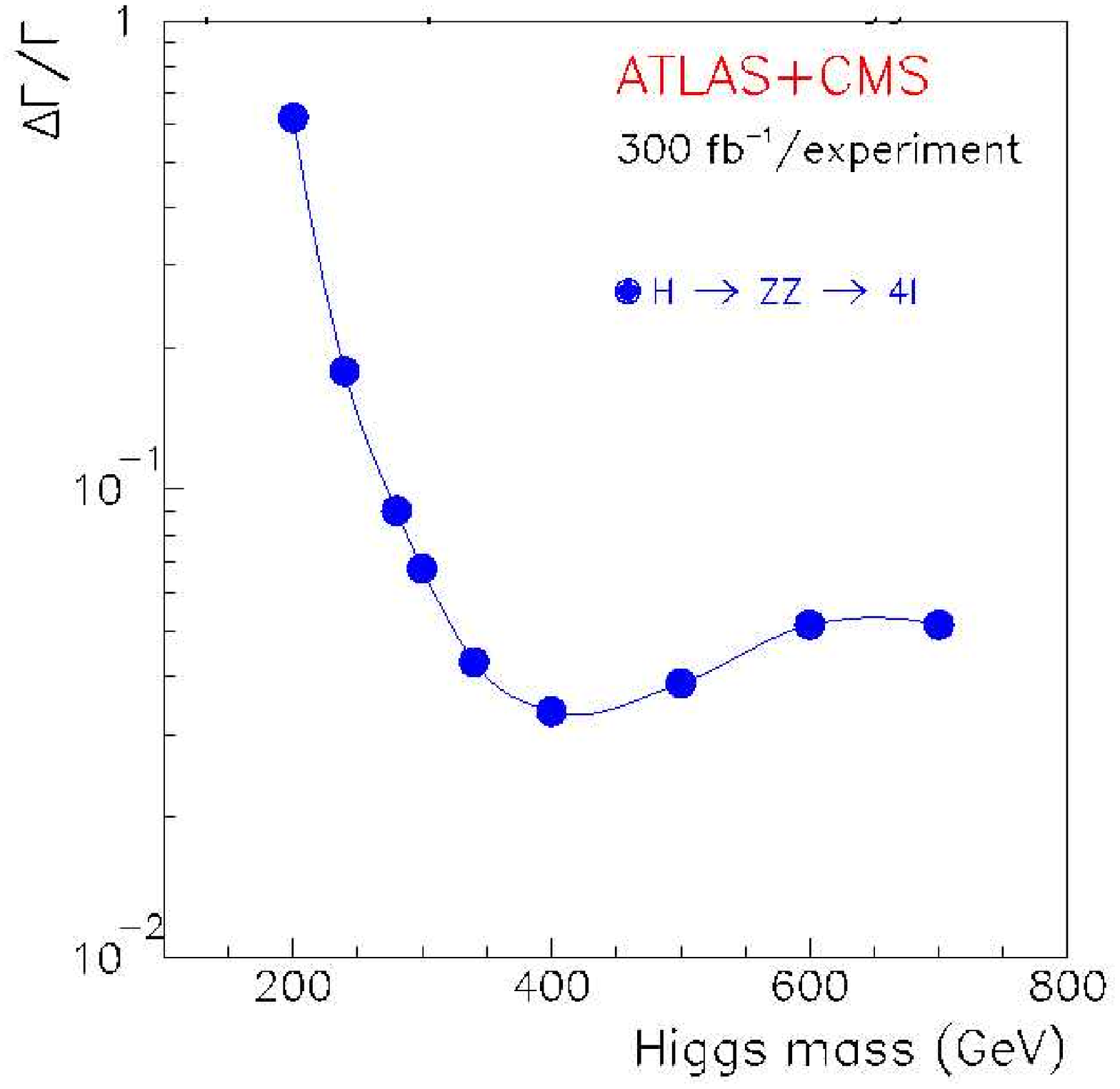}
}
\caption{\em Expected  accuracy of the measurement of Higgs mass and 
width at LHC\protect\cite{fabiola}.\label{fthree}}
\end{figure}
fig.~\ref{fthree} shows that at lower values of $m_h$,  simultaneous use  of
different channels is very useful. For $m_h \gsim 500$ GeV a better theoretical
analysis of the resonant signal and the nonresonant background is still 
lacking. As far as the width $\Gamma_h$ is concerned, a measurement is possible
only for $m_h > 200$ GeV, at a level of $\sim 5\%$ and that too
at the end of three years of the high luminosity run. At this value of $m_h$ 
the information essentially comes from the $h \to ZZ \to 4 l$ 
channel. The values of
$\Delta \Gamma / \Gamma$ that can be reached at the end of three years of high
luminosity run, obtained in a combined ATLAS and CMS study, are shown in 
the figure in the right panel in fig.~\ref{fthree}. 

Apart from the precision measurements of the mass and the width of the 
Higgs particle, possible accuracy of extraction of the couplings of the
Higgs with the matter and gauge particles, with a view to  check the 
spontaneous symmetry breaking scenario, is also an important issue. 
\begin{table}[htb]
\caption{\em Expected accuracy in the extraction of the Higgs couplings as
evaluated by ATLAS\protect\cite{fabiola}.\label{tab:tab2}}
\begin{center}
\begin{tabular}{|c|c|c|}
\hline
Ratio of cross-sections&Ratio of extracted &Expected Accuracy\\
&Couplings&Mass Range\\
\hline
&&\\
$\frac{\sigma (t \bar t h + Wh)(h \to \gamma \gamma)}{\sigma (t \bar t h + Wh)(h \to b \bar b)}$&$\frac{B.R.(h \to \gamma \gamma)}{B.R.(h \to b  \bar b)}$&$\sim 15\%$, 80-120 GeV\\
&&\\
\hline
&&\\
$\frac{\sigma (h \to \gamma \gamma)}{\sigma (h \to 4 l)}$&
$\frac{B.R.(h \to \gamma \gamma)}{B.R.(h \to Z Z^{(*)})}$&$\sim 7\%$, 
120-150 GeV\\
&&\\
\hline
&&\\
$\frac{\sigma (t \bar t h \to \gamma \gamma/b \bar b)}
{\sigma ( Wh \to \gamma \gamma /b \bar b)}$& $\frac{g^2_{h t \bar t}}{g^2_{hWW}}$&$\sim 15\%, 80 < m_h < 120$ GeV\\
&&\\
\hline
&&\\
$\frac{\sigma (h \to Z Z^*\to 4 l)}{\sigma ( h \to W W^* \to l \nu l \nu)}$& 
$\frac{g^2_{hZZ}}{g^2_{hWW}}$&$\sim 10\%, 130 < m_h < 190$ GeV\\
&&\\
\hline
\end{tabular}
\end{center}
\end{table}
Table~\ref{tab:tab2} shows the accuracy which would be possible in 
extracting ratios of various couplings, according to an analysis by 
the ATLAS collaboration. This analysis is done by measuring the 
ratios of cross-sections so that the measurement is insensitive to 
the theoretical uncertainties in the prediction of hadronic 
cross-sections. All these measurements use only the inlclusive 
Higgs mode. 

New analyses based on an idea by Zeppenfeld and 
collaborators\cite{Dieter-kaoru,plehn,dieter-mad,richter-was} have explored the
use of production of the Higgs via WW/ZZ(IVB) fusion, in the process
$pp \to q + q + V + V + X \to q + q + h + X.$ Here the two jets go in the 
forward direction. The observation\cite{Dieter-kaoru} that the colour flow for
the IVB fusion production process is quite different than the background,
suggested the use of veto against jets in the central region, to enhance the Higgs
signal produced via the IVB fusion. This has increased the possibility
of studying the Higgs production via IVB fusion process to lower values
of $m_h (< 120 {\rm GeV})$ than previously thought possible. E.g., a parton
level Monte Carlo\cite{plehn} has demonstrated that already for 
$\int {\cal L} dt = 30  fb^{-1}, qq \to qqh \to qq \tau^+ \tau^-$
gives $S/B = 2/1$. This is to be contrasted with the luminosities required
for a similar level of significance, in the inclusive channel shown in 
fig.~\ref{ftwo}. It has been demonstrated\cite{dieter-mad} that using the 
production  of Higgs in the process $qq \to h jj$, followed by the decay
of the Higgs into various channels $\gamma \gamma, \tau^+ \tau^-, W^+ W^-$ 
as well as the inclusive channels $gg \to h \gamma \gamma, gg \to h 
\to Z Z^{(*)}$, it should be possible to measure $\Gamma_h, g_{hff} $ 
and $g_{hWW}$ to a level of $10-20 \%$ , assuming
that $\Gamma (h \to b \bar b) / \Gamma (h \to \tau^+ \tau^-)$ has approximately
the SM value. Recall, here that after a full LHC run,with a combined CMS +ATLAS
analysis, the latter should be known to $\sim 15 \%$. This strategy is being 
studied further at the detector level\cite{richter-was}.
In principle, such measurements of the Higgs couplings might be an indirect way
to look for the effect of physics beyond the SM. We will discuss this later.

By the start of the LHC (unless the LEP has already confirmed the `light' higgs,
by the time these words appear in print!), with the possible TeV 33 run with
$\int {\cal L} dt = 30  fb^{-1}$, Tevatron can give us an indication and 
a possible signal for a light Higgs, combining the information from 
different  associated production  modes: $Wh,Zh $ and $W W^{(*)}$. The 
inclusive channel $\gamma \gamma /4l$  which will be dominantly used at LHC is 
completely useless at Tevatron. So in  some sense the imformation we get from 
Tevtaron/LHC will be complementary. 

Thus in summary the LHC, after one year of operation should be able to see 
the SM higgs if it is in the mass range where the SM says it should be. Further
at the end of $\sim 6$ years the ratio of various couplings of $h$ will be 
known 
within $\sim 10 \%$. One example of the physics beyond the SM that such a measurement
will probe indirectly is supersymmetry (SUSY). The direct manifestation
of SUSY for the Higgs sector will of course be the presence of the extra 
scalars expected in the MSSM.

\section{Search at LHC for the MSSM higgs}
As is well known the MSSM Higgs sector is much richer and has five scalars;
three neutrals: CP even \sh\ , H and CP odd A as well as the pair of charged
Higgses $H^\pm$. So many more search channels are available. The most important 
aspect of the MSSM higgs, however is the upper limit\cite{heinemeyer,quiros}
 of $130$ GeV ($200$ GeV) for MSSM (and its extensions), on
the mass of lightest higgs. The masses and couplings of these scalars depend
on the supersymmetric parameters $m_A, \tan \beta$ as well as SUSY breaking 
parameters $m_{\tilde t_1}$ and the mixing in the stop sector controlled
essentially by $A_t$. In general the couplings of the \sh\ can be quite 
different from the SM higgs $h$; {\it e.g.} even for large 
$m_A ( > 400 {\rm GeV}), \Gamma_{\sh} /\Gamma_{h} > 0.8,$ over most
of the range of all the other parameters. Thus such measurements can be 
a `harbinger' of SUSY. The upper limit on the mass of \sh\ forbids its
decays into a $VV$ pair and thus it  is much narrower than the SM $h$. Hence
the only decays that can be employed for the search of \sh\ are $b \bar b,
\gamma \gamma$ and $\tau^+ \tau^- $. The last can be exploited mainly in 
the large $\tan \beta$ range. This can be perhaps be used even  more
effectively using the $qq \sh$ mode\cite{plehn}.  The $\gamma \gamma$ mode
can be suppressed  for the \sh\ compared to $h$.  
The reduction is substantial even when all the sparticles are heavy,
at low $m_A, \tan \beta$. 
\begin{figure}[htb]
\begin{center}
\includegraphics*[scale=0.4]{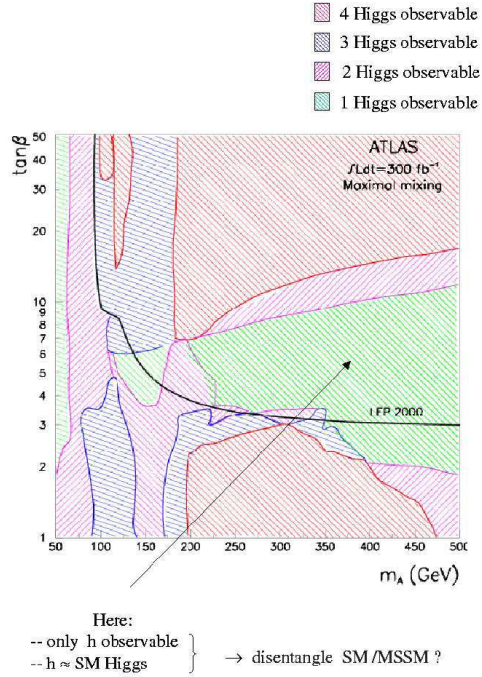}
\caption{\em  Number of MSSM scalars observable at LHC  in different regions of
$\tan \beta - M_A$ plane\protect\cite{fabiola}.}
\label{ffour}
\end{center}
\end{figure}
Fig.~\ref{ffour} shows  various regions in the $\tan \beta - m_A$ plane
divided according to the number of the MSSM scalars observable at LHC, 
according to an ATLAS analysis, for the case of maximal mixing in the stop 
sector, at the end of full LHC run. This shows that for low $M_A (\gsim
200$ GeV) and low $\tan \beta (\lessapprox 8-9)$, only one of the five MSSM 
scalars will be observable. Recall  further that the differences in the 
coupling of \sh\ and $h$ are quite small in this region.

Situation can be considerably worse if some of the sparticles, particularly
$\tilde t$ and $\tilde \chi_i^{\pm}, \tilde \chi_i^{0}$ are light. Light 
stop/charginos can decrease $\Gamma (\sh \to \gamma \gamma)$ 
through their contribution in the loop. For the light $\tilde t$ the inclusive 
production mode $gg \to \sh$ is also reduced substantially. If the channel 
$\sh \to \chi^0_1 \chi^0_1$ is open, that depresses the $BR$ into the 
$\gamma \gamma$ channel even further\cite{abdel1,abdel2,bbs,bbdgr}.
\begin{figure}[htb]
        \centerline{
             \includegraphics*[scale=0.50]{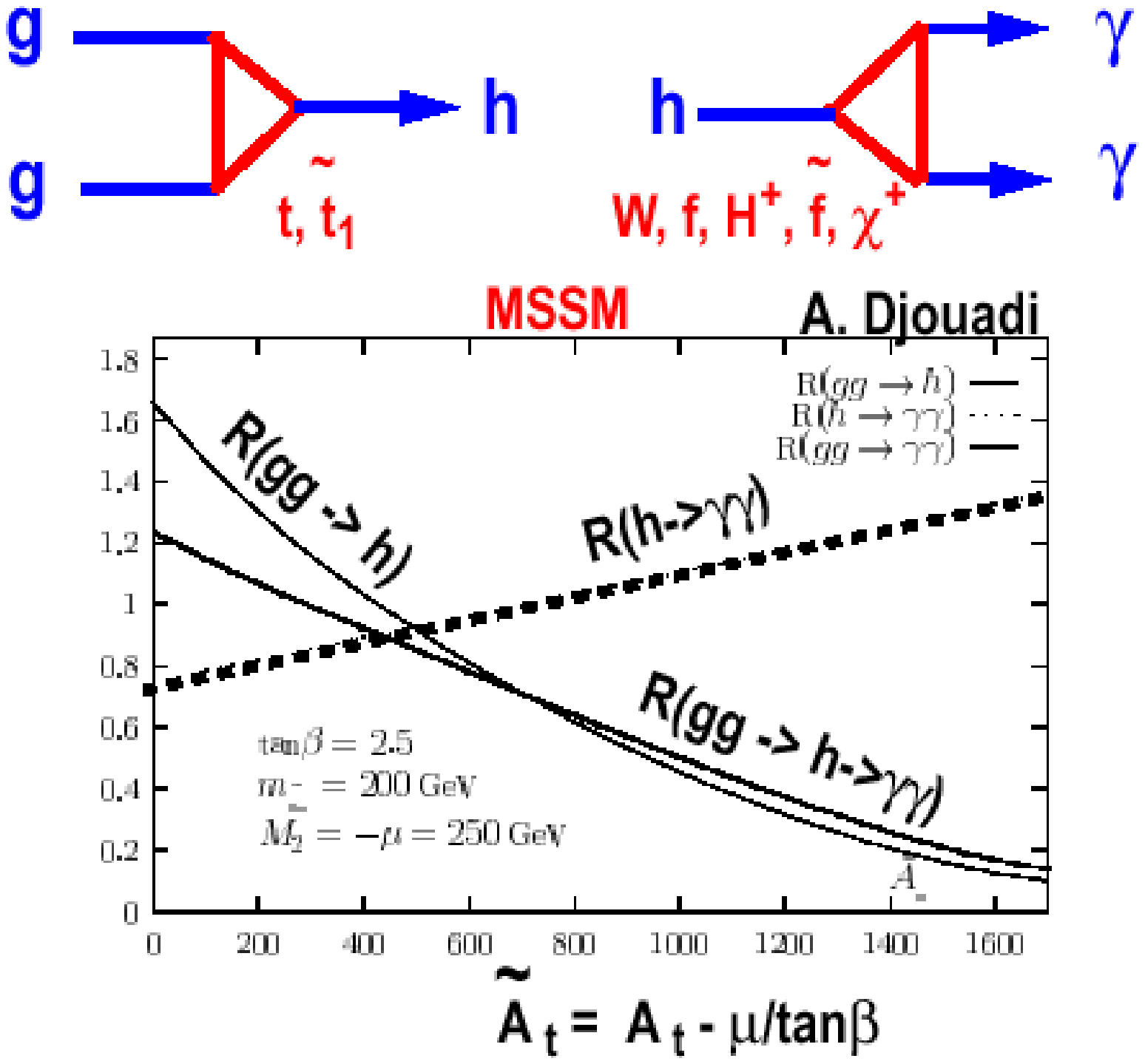}
              \includegraphics*[scale=0.30]{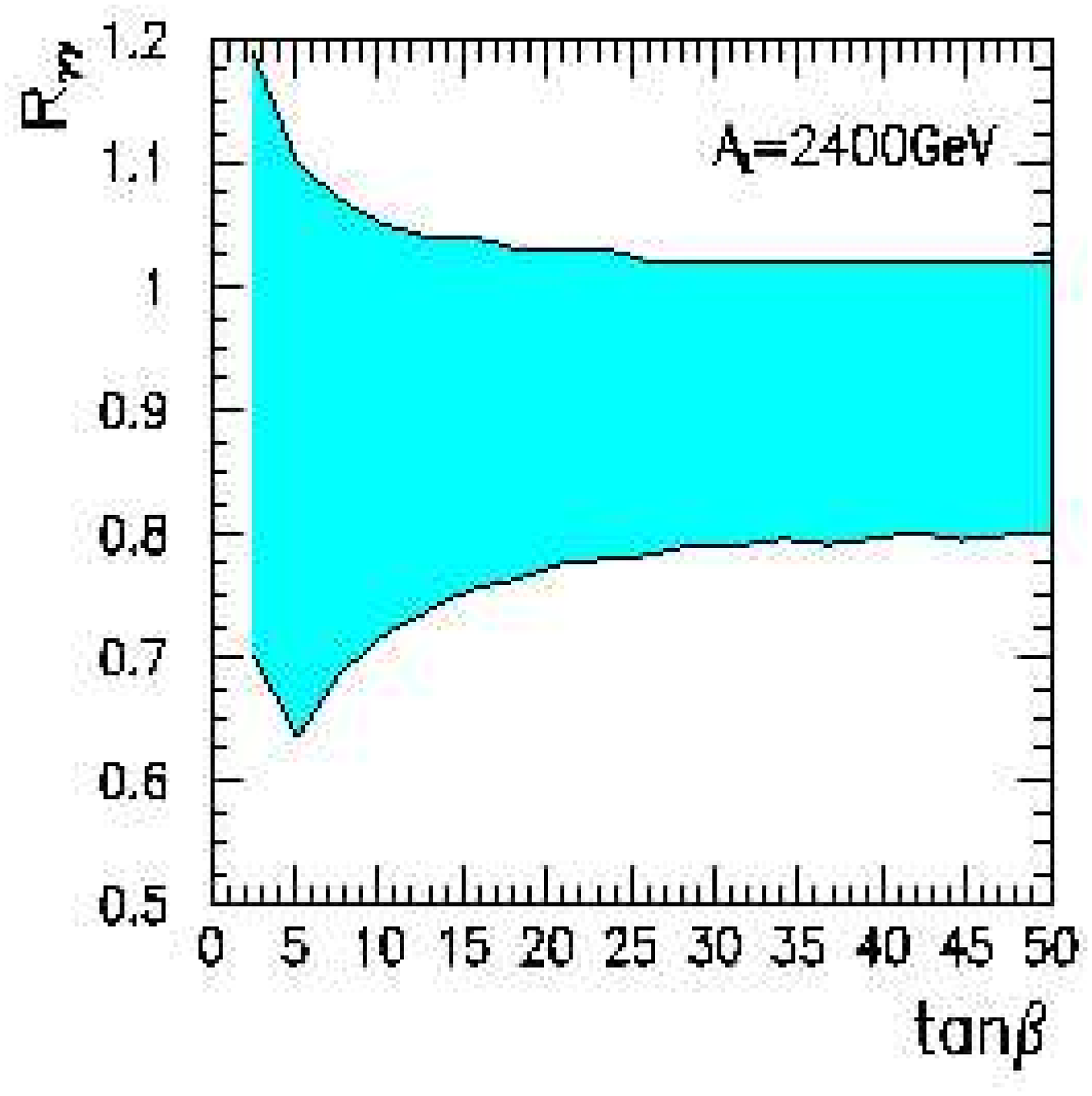}
}
\caption{\em  Effect of light sparticles on the $\gamma \gamma$ decay
width  and $gg$ production of the Higgs\protect\cite{abdel1,bbdgr}.
\label{ffive}}
\end{figure}
The left panel in fig.~\ref{ffive}\cite{abdel1} shows the ratio
$$R(h \to \gamma \gamma) = \frac{\Gamma (\sh \to \gamma \gamma)}
{\Gamma (h \to \gamma \gamma)}$$
and ratios $R(gg \to h), R(gg \to h \gamma \gamma)$ defined similarly.  Thus
we see that for low $\tan \beta$ the signal for the light neutral higgs
\sh\ can be completely lost for a light stop.  
The panel on the right in fig.~\ref{ffive}\cite{bbdgr}
shows $R(\sh \to \gamma \gamma)$ as a function of $\tan \beta$ for the case of
only a light chargino and neutralino. Luckily, eventhough light sparticles, 
particularly a light $\tilde t$ can cause disappearance of this signal, 
associated production of the  higgs \sh\ in the channel ${\tilde t}_1 
{\tilde t}^*_1 \sh /t \bar t \sh$ provides a viable signal. However, 
an analysis of
the optimisation of the observability of such a light stop $(m_{\tilde t} 
\simeq 100-200 $ GeV) at the LHC still needs to be done. This brings
us to the subject of search for Supersymmetry at the LHC.

\section{Prospects for  SUSY search at the LHC}
The new developements in the past years in the subject have been in trying
to set up strategies so as to disentangle signals due to different 
sparticles from each other and extract information about the SUSY breaking
scale and mechanism, from the experimentally determined properties and the 
spectrum of the sparticles. As we know, the couplings of  {\bf almost} all
the sparticles are determined by the symmetry, except for the charginos,
neutralinos and the light $\tilde t$. However, masses of all
the sparticles are completely model dependent. The four different types of
models that are normally considered are
\begin{enumerate}
\item (M)SUGRA: The highly constrained supergravity model which is
characterised by just five parameters,
\item (C)MSSM: MSSM where the number of parameters is reduced by some 
very reasonable assumptions,
\item AMSB: Models which have Anamoly mediated SUSY breaking,
\item GMSB: Models which have gauge mediated SUSY breaking.
\end{enumerate}
In cases 1-3 SUSY is broken via gravitational effects and 
${\tilde \chi}_1^0$ is the Lightest Supersymmetric Particle(LSP). 
In case 4 the  LSP is the light gravitino and the phenomenology is 
decided by the life time of the next lightest superymmetric particle 
(NLSP) which can be either a $\tilde \tau$ or ${\tilde \chi}_1^0$.
In cases 1-3 the missing transverse energy \et\ is the main signal.
In case 4, along with \et\ the final states also have photons
and/or displaced vertices, stable charged particle tracks etc. as
the telltale signals of SUSY. 
\begin{figure}[htb]
\begin{center}
\includegraphics*[scale=0.4]{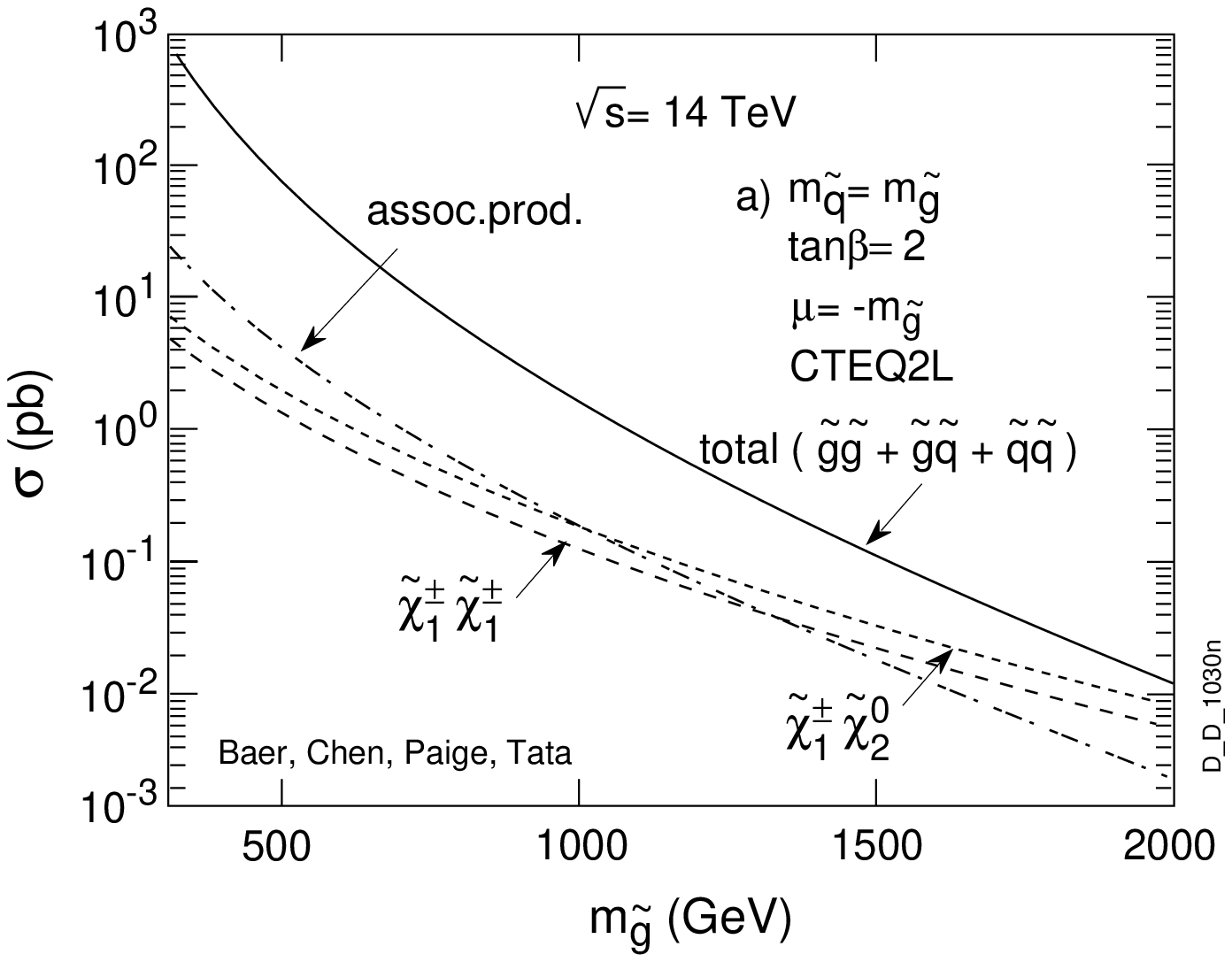}
\caption{\em Expected production cross-sections for various sparticles at
the LHC\protect\cite{tata}.}
\label{fsix}
\end{center}
\end{figure}
As is clear from the fig.~\ref{fsix} LHC is best suited for the search
of the strongly interacting $\tilde g, \tilde q$ because they have the 
strongest production rates.  The  ${\tilde\chi}_i^{\pm}, {\tilde \chi}^0_i$,
are produced via the EW processes or the decays of the $\tilde g, \tilde q$.
The former mode of production gives very clear signal of  `hadronically
quiet' events. The sleptons which can be produced mainly via the DY
process have the smallest cross-section. As mentioned earlier, various
sparticles can give rise to similar final states, depending on the mass
hierarchy. Thus, at LHC the most complicated background to SUSY search is
SUSY itself! The signals consist of events with \et, $m$ leptons and
$n$ jets with $m,n \geq 0$. Most of the detailed simulations which address the issue of the
reach of LHC for SUSY scale, have been done in the context of (M)SUGRA
picture. 
\begin{figure}[htb]
\begin{center}
\epsfxsize=12cm\epsfysize=10.0cm
                     \epsfbox{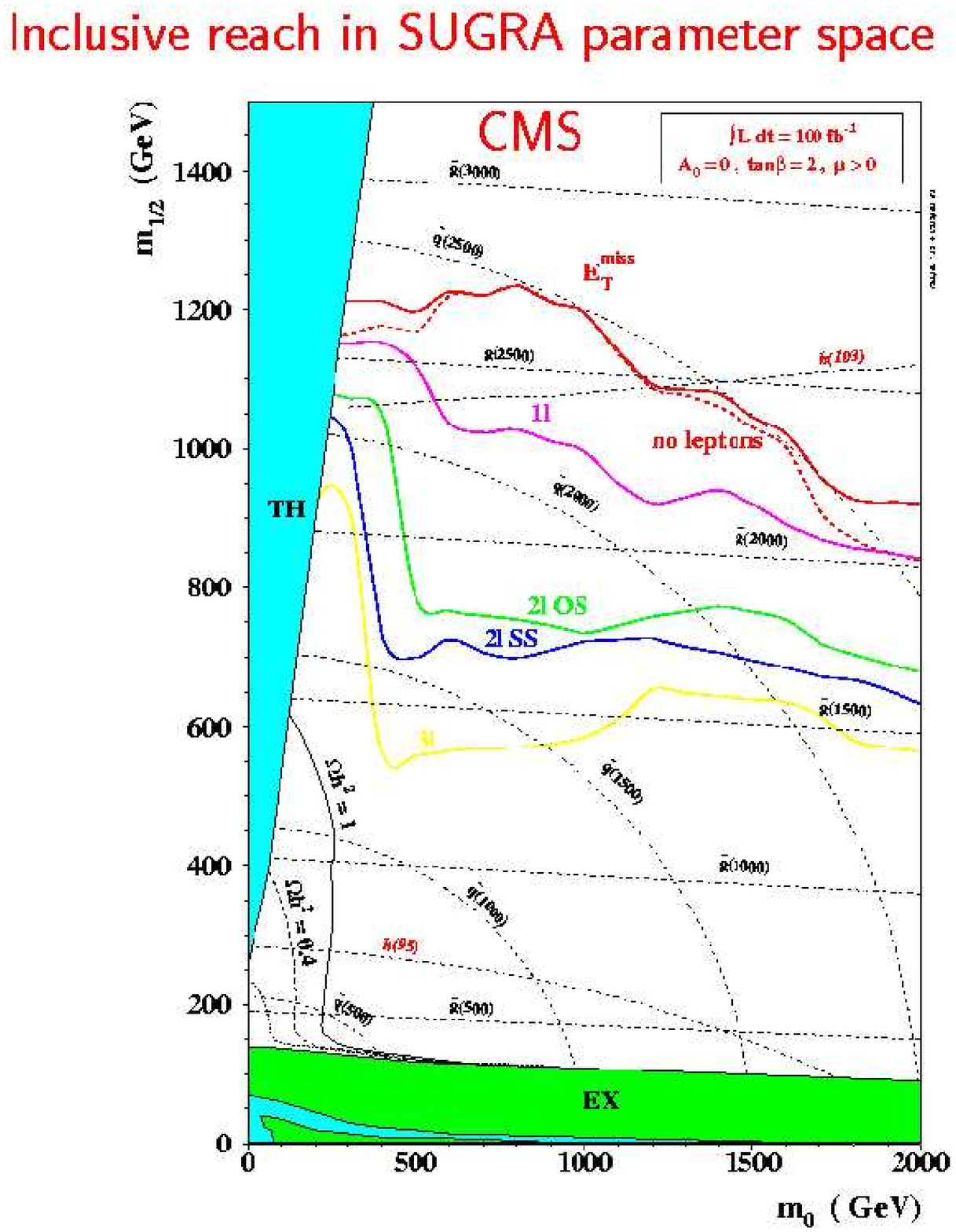}
\caption{\em Expected reach for SUSY searches at the LHC\protect\cite{polsel}.\label{fseven}}
\end{center}
\end{figure}
We see from fig.\ref{fseven} that for $\tilde g, \tilde q$ the
reach at LHC is about $2.5$ TeV and over most of the parameter space 
multiple signals are observable.

To determine the SUSY breaking scale $M_{SUSY}$ from the jet events, a method 
suggested by Hinchliffe et al\cite{paige} is used, which consists in defining
$$M_{eff}  =  \sum_{i=1}^4 |P_{T(i)}| + \et$$ 
and look at the distribution in $M_{eff}$. The jets, that are produced by 
sparticle production and decay, will have
$P_T \propto m_1 - \frac{m_2^2}{m_1}$, where $m_1,m_2$ are the masses of the
decaying sparticles. Thus this distribution can be used to determine 
$M_{SUSY}$.
\begin{figure}[htb]
\begin{center}
\includegraphics*[scale=0.4]{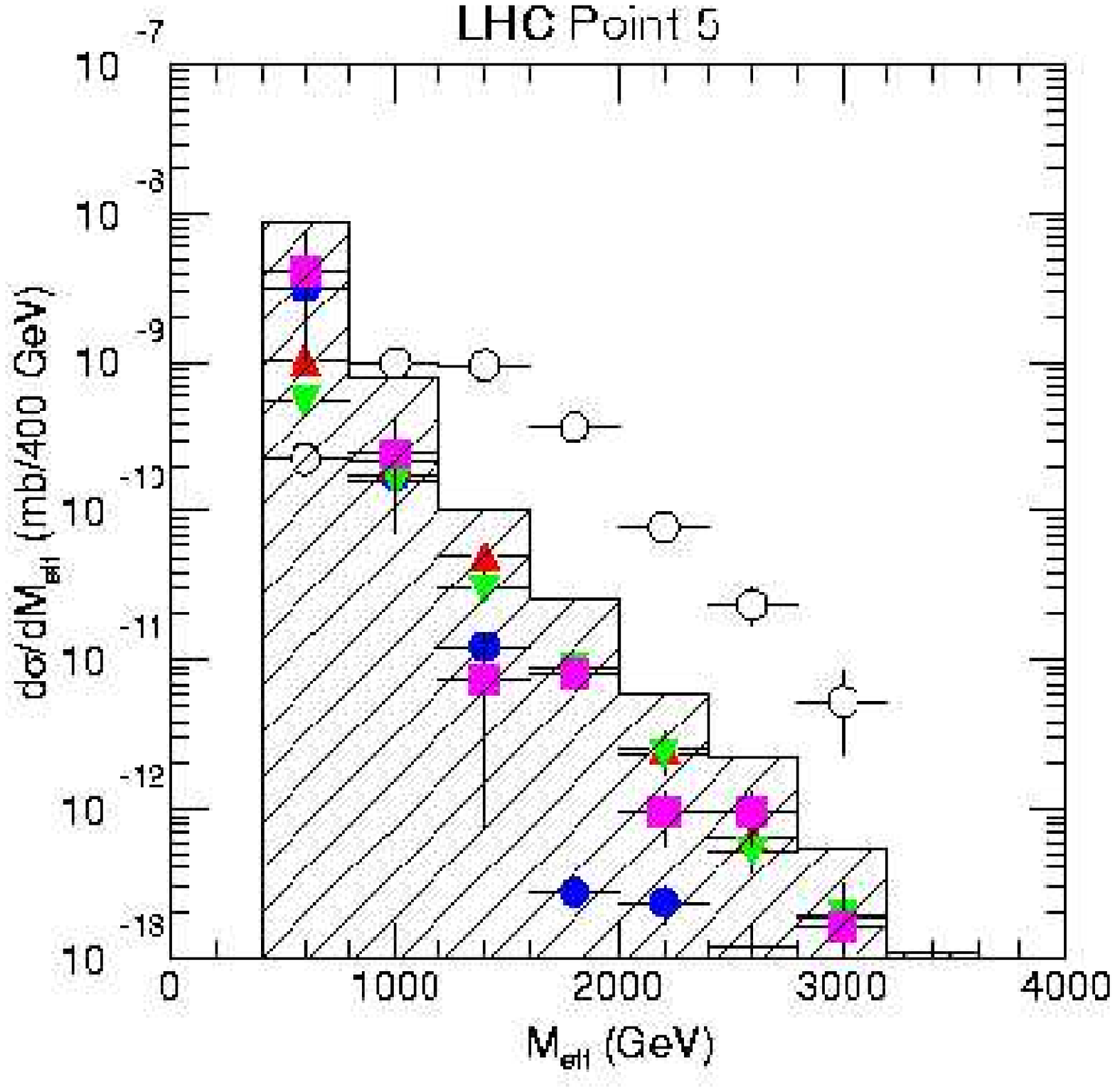}
\caption{\em  Determination of SUSY breaking scale using jet events at 
LHC point \protect\cite{paige}.\label{feight}}
\end{center}
\end{figure}
The distribution in fig.~\ref{feight} shows that indeed there is 
a shoulder above the SM background. The scale $M_{SUSY}$ is defined either 
from the peak position or the point where the signal is aprroximately equal to the
background. Then of course one checks how well $M_{SUSY}$ so determined
tracks the input scale. A high degree of correlation was observed in
the analysis, implying that this can be a way to determine the SUSY breaking 
scale in a precise manner. A recent analysis\cite{tovey} shows that 
while for (M)SUGRA and (C)MSSM a precision of $\sim 0.3 \% $ and $\sim 3\%$ can be 
reached, albeit for a very high integrated luminosity of about 
$1000  fb^{-1}$, for GMSB models the accuracy is only about
$20\%$. 

It is possible to reconstruct the masses of the charginos/neutralinos
using kinematic distributions.
\begin{figure}[htb]
\begin{center}
\includegraphics*[scale=0.4]{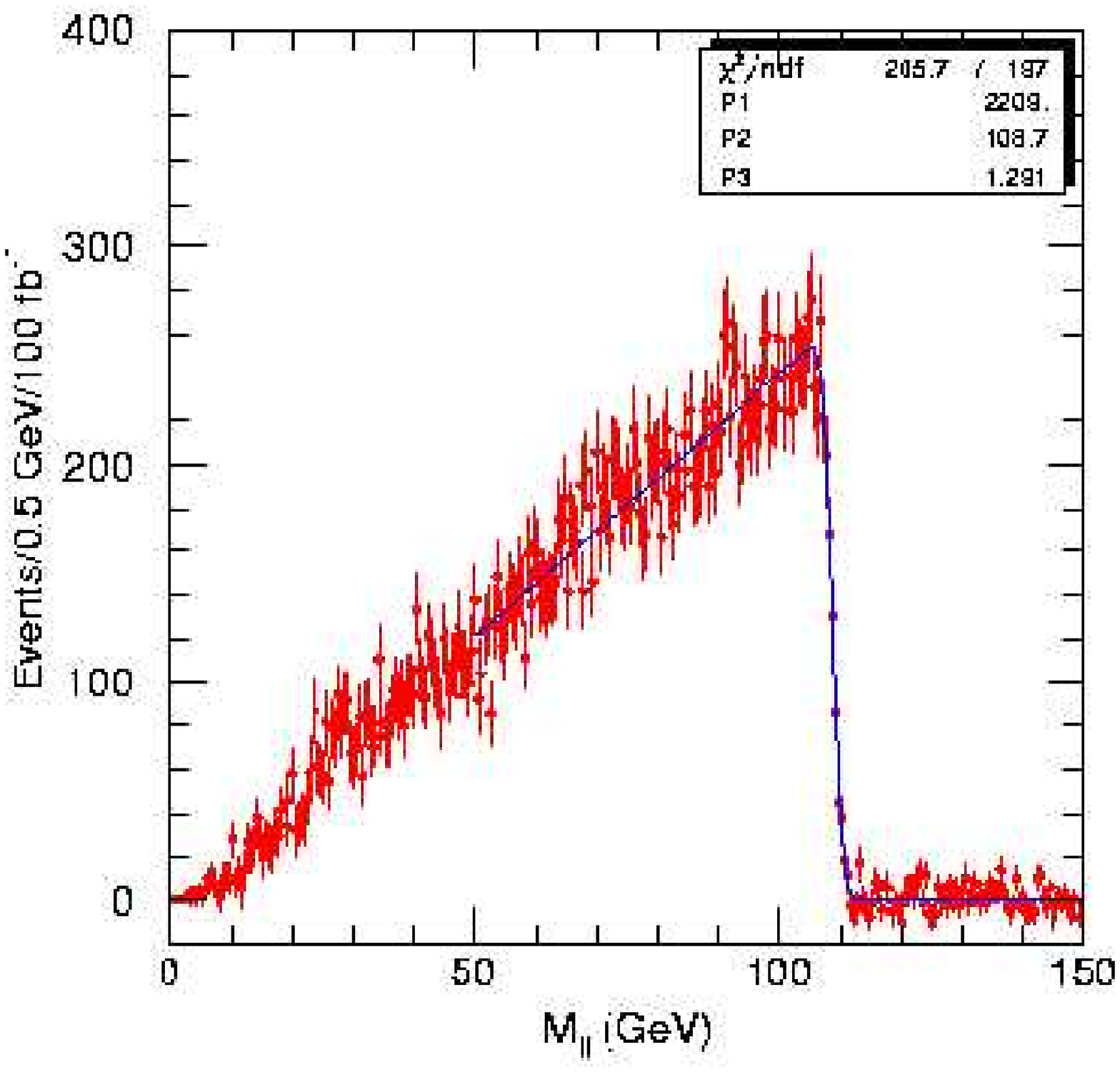}
\caption{\em Kinematic reconstruction of the mass of ${\tilde\chi}_2^0$ 
from the dilepton mass distribution~\protect\cite{polsel}.}
\label{fnine}
\end{center}
\end{figure}
Fig.~\ref{fnine} demonstrates this, using the distribution in the
invariant mass $m_{l^+l^-}$ for the $l^+l^-$ pair produced in the
decay ${\tilde \chi}^0_2 \to {\tilde \chi}_1^0 l^+ l^-$. The end 
point of this distribution is $\sim m_{{\tilde\chi}_2^0} - m_{{\tilde \chi}_1^0}$. However, such analyses have to be performed with caution. As 
pointed out by
Nojiri et al\cite{mihoko1}, the shape of the spectrum near the end point can
at times depend very strongly on the dynamics such as the composition of the
neutralino and the slepton mass. One can still use these determinations
to extract model parameters, but one has to be careful.

\section{`Large' extra dimensions at LHC}
The whole development of the subject of `large' extra dimensions at LHC
is a very good example as to  how the various features of the 
detectors, 
such as good lepton detection, can be used very effectively in looking
for `new' physics which was {\bf not} taken into account while designing
the detector. There have been a lot of discussions on the subject and it
will be reviewd\cite{peskin} in the proceedings elsewhere. In the context of
LHC, the clearest signal for these `large' dimensions is via the observation
of graviton resonances\cite{hewett} in the dilepton spectrum via the process
$gg \to G \to l^+l^-$. It has been demonstrated by Hewett et al\cite{hewett} 
that by using the constraints already available from the dijet/dilepton
data from the Tevatron and making reasonable assumptions so that the EW scale
is free from hierarchy problem, in the scenario with `warped' extra 
dimesnions\cite{RS}, the parameter space of the model can be completely 
covered at LHC using the dilepton channel.

Apart from determining the mass of the graviton, it is also essntial to 
check the spin of the exchanged particle. ATLAS performed an 
analysis\cite{park}, which showed that the acceptance of the detector 
is quite low at large $\cos \theta^*$. 
\begin{figure}[htb]
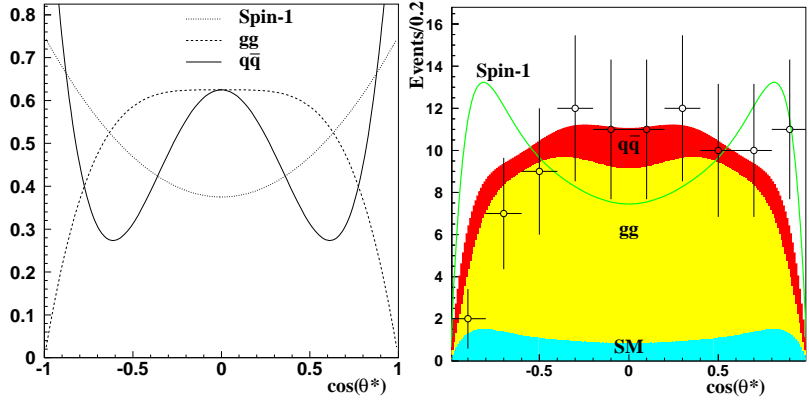

      \centerline{
      \includegraphics*[scale=0.3]{ang_theory.epsi}
      \includegraphics*[scale=0.3]{ang_experiment_1500.epsi}
  }
\caption{\em Angular distribution in the cm. frame for the $l^+l^-$ 
pair expected in the detector for the graviton along with the  expectation 
for a spin 1 particle at LHC\protect\cite{park}.}
\label{ften}
\end{figure}
The left panel of fig.~\ref{ften} shows the different angular distributions expected for
different spin exchanges. For the spin-2 case the contributions from the
$gg$ and $qq$ initial state are shown separately. The panel on the right 
shows that even with the lowered acceptance, it might be possible to 
discriminate against a spin one case.
Many more investigations on the subject are going on and the  end
conclusion is that it should be possible to see the effect of these
`large' compact dimensions (warped or otherwise) at the LHC upto 
almost all the values of the model parameters which seem reasonable and 
for which the theoretical formulation remains consistent.

\section{Conlcusions}
\begin{description}
\item[1)] LHC is capable of finding the SM higgs with a high level of 
significance over the entire mass range implied by the SM, at the end
of one year of running. 
\item[2)] The mass of the higgs can be measured at $\sim 0.1\%$ level and the
width (for $m_h > 200 $ GeV) at $\sim 5\%$ level after a total of five years of
LHC running. The current analyses also indicate that at the end of three years 
of high luminosity run, the couplings  can also be extracted to about
$10-20\%$ level even for a light higgs.
\item[3)]LHC is capable of covering the entire MSSM parameter space for the 
Higgs search with $300  fb^{-1}$ luminosity (at the end of five years 
in all), but the search in the $\gamma \gamma$ channel {\it may} not 
be always  possible or the lightest neutral \sh\ may not always
be observable. This can happen due to the effect of light sparticles, mainly 
a light $\tilde t$ with $m_{\tilde t} \simeq 100-200 $ GeV. Search strategies 
for such a light  stop  at LHC need to be optimised.
\item[4)] $\tilde q,\tilde g$ can be discovered if  they are lighter than 
$\lessapprox 2.5$ TeV; the sleptons, if lighter than $\lessapprox 340$ GeV and
the charginos/neutralinos if lighter than $\lessapprox 500-600$ GeV.
All these estimates of the  limits have been obtained in analyses which 
asume a (M)SUGRA scenario. Similar analyses for the GMSB/AMSB scenarios
are underway. Further in the framework of (M)SUGRA and (C)MSSM, a precise
determination of model parameters and hence of SUSY breaking scale,
is possible from kinematic recosntructions of 
sparticle masses. However, for GMSB the analysis has not been optimised
yet and the reconstruction of SUSY scale seems possible only 
at the level of $20\%$. This would require $\int{\cal L} dt = 
1000  fb^{-1}$.
\item[5)] The RS scenario of `large' extra dimensions can be confirmed 
or ruled out over all the reasonable range of model parameters by LHC.
\end{description}

\section*{Acknowledgments}
I wish to thank the organisers of the APPC 2000 and III ACFA LC Workshop
for organising an excellent meeting and providing a very nice atmosphere
for discussions.


\begin{thebibliography}{99}
\bibitem{Komamiya} S. Komamiya, {\em Talk in these proceedings}.
\bibitem{hambye} T. Hambye and K. Riesselmann,
\Journal{\PRD}{55}{7255}{1997};{\bf hep-ph/9708416} in 
{\em ECFA/DESY study on particles and detectors for 
for the linear colliders}, Ed. R. Settlers, {\bf DESY 97-123E}.
\bibitem{heinemeyer} S. Heinemeyer, W. Hollik and G. Weiglein,
\Journal{\PLB}{455}{179}{1999}; \Journal{\EUJ} {9}{343}{1999};
R. J. Zhang, \Journal{PLB}{447}{89}{1999}, J.R. Espinoza and 
R.J. Zhang, \Journal{JHEP}{3}{26}{2000},{\bf hep-ph/0003246};
M. Carena, H.E. Haber, S. Heinemeyer, W. Hollik, C.E.M. Wagner and G. Weiglein,
\Journal{\NPB}{580}{29}{2000}.
\bibitem{quiros} J.R. Espinoza and M. Quiros, \Journal{\PRL}{81}{516}{1998}.
\bibitem{barbie} R. Barbieri and A. Strumia, \Journal{\PLB}{462}{144}{1999},
{\bf hep-ph/0005203}, {\bf hep-ph/0007265}.
\bibitem{murayama} H. Murayama and C. Kolda, \Journal{\JHEP}{7}{35}{2000}.
\bibitem{ADD} N. Arkani-Hamed, S. Dimoupoulos and G. Dvali, 
\Journal{\PLB}{429}{263}{1998},\Journal{\PRD}{59}{086004}{1999};
I. Antoniadis, N. Arkani-Hamed, S. Dimopoulous and G. Dvali, 
\Journal{\PLB}{436}{257}{1998}.
\bibitem{RS}L. Randall and R. Sundrum, \Journal{\PRL}{83}{3370}
{1999}, {\it ibid.}, 4690 (1999).
\bibitem{bhat} P.C. Bhat, R. Gilmartin and H.B. Prosper,
{\em Fermilab-Pub-00/006} \Journal{\PRD}{62}{074022}{2000}.
\bibitem{michael} M. Dittmar, \Journal{\pramana}{55}{151}{2000}.
\bibitem{Binoth} T. Bionth, {\bf hep-ph/0005194}; T. Binoth, J.P. Guillet, 
E. Pillon and M. Werlen, \Journal{\EUJ}{16}{311}{2000}.
\bibitem{kunszt} D. de Florian and Z. Kunszt, \Journal{\PLB}{460}{184}{1999}.
\bibitem{Abdullin} S. Abdullin et al., {\bf hep-ph/9805341}.
\bibitem{Djouadi} C. Balazs, A. Djouadi, V. Ilyin and M. Spira, 
{\em in the Higgs Working Group for the workshop `Physics at the 
TeV colliders' Les Houches, June 1999}, {\bf hep-ph/0002258}.
\bibitem{dreiner} M. Dittmar and H. Dreiner, \Journal{\PRD}{55}
{167}{1997}.
\bibitem{fabiola} F. Gianotti, Talk presented  at LHCC meeting,
http://gianotti.home.cern.ch/gianotti/phys\_info.html, F. Gianotti and 
M. Pepe Altarelli, {\bf hep-ph/0006016}.
\bibitem{Dieter-kaoru} D. Rainwater,  K. Hagiwara and D. Zeppenfeld, 
\Journal{\PRD}{59}{014037}{1999}.
\bibitem{plehn} T. Plehn, D. Rainwater and D. Zeppenfeld, \Journal{\PRD}{61}
{093005}{2000}.
\bibitem{dieter-mad} D. Zeppenfeld, {\bf hep-ph 0005151}.
\bibitem{richter-was} D. Zeppenfeld, R. Kinnunen, A. Nikitenko and E. Richter-W{\c{a}}s, \Journal{\PRD}{62}{013009}{2000}.
\bibitem{abdel1} A. Djouadi, \Journal{\PLB}{435}{1998}{101}.
\bibitem{abdel2} A. Djouadi, {\bf hep-ph 9903382}. 
\bibitem{bbs} G.Belanger, F. Boudjema and K. Sridhar,
\Journal{\NPB}{568}{3}{2000}.
\bibitem{bbdgr} G. Belanger, F. Boudjema, F. Donato, R. Godbole and
S. Rosier-Lees, \Journal{\NPB}{581}{3}{2000}.
\bibitem{tata} H. Baer, C. Chen, F. Paige and X. Tata, \Journal{\PRD}{53}{6241}{1996}.
\bibitem{polsel} G. Polsello, {\em Talk presented at the SUSY2K, June 2000},
http://wwwth.cern.ch/susy2k/susy2kfinalprog.html.
\bibitem{paige} I. Hinchliffe, F.E. Paige, M.D. Shapiro, J. Soderqvist and W. 
Yao, \Journal{\PRD}{55}{5520}{1997}.
\bibitem{tovey} D.R. Tovey, {\bf hep-ph/0006276}.
\bibitem{mihoko1}M. M. Nojiri and Y. Yamada, \Journal{\PRD}{60}{015006}{1999}.
\bibitem{peskin} M.E. Peskin, {\em Talk in these proceedings}.
\bibitem{hewett} H. Davoudiasl, J.L. Hewett and T.Rizzo, 
\Journal{\PLB}{473} {49}{2000}, \Journal{\PRL}{84}{2080}{2000}, 
{\bf hep-ph/0006041}.
\bibitem{park} B. C. Allanach, K. Odagiri, M.A. Parker and B.R. Webber, 
\Journal{\JHEP}{9}{019}{2000}.
\end{thebibliography}
\end{document}